# How Does Kanban Impact Communication and Collaboration in Software Engineering Teams?

Nilay Oza[1], Fabian Fagerholm[2], Jürgen Münch[3]
University of Helsinki
P.O. Box 68 (Gustaf Hällströminkatu 2b)
FI-00014, Finland
nilay.oza@cs.helsinki.fi[1], fabian.fagerholm@helsinki.fi[2], juergen.muench@cs.helsinki.fi[3]

*Abstract*— Highly iterative development processes such as Kanban have gained significant importance in industry. However, the impact of such processes on team collaboration and communication is widely unknown. In this paper, we analyze how the Kanban process aids software team's behaviors – in particular, communication and collaboration. The team under study developed a mobile payment software product in six iterations over seven weeks. The data were collected by a questionnaire, repeated at the end of each iteration. The results indicate that Kanban has a positive effect at the beginning to get the team working together to identify and coordinate the work. Later phases, when the team members have established good rapport among them, the importance for facilitating team collaboration could not be shown. Results also indicate that Kanban helps team members to collectively identify and surface the missing tasks to keep the pace of the development harmonized across the whole team, resulting into increased collaboration. Besides presenting the study and the results, the article gives an outlook on future work.

## I. Introduction

It is already well established that team communication and collaboration are critical to the success of software development [1, 6]. As early as 1979, Basili [8] highlighted the importance of investigation into human factors in software development to induce measureable differences in both the development process and the developed product. Since then, new ways of software development have constantly influenced how teams communicate and collaborate. For example, agile methodologies have led to significant increases in interactions between members of a team [6, 7]. The software industry has been increasingly using Kanban in managing software development projects [2]. In this paper, we present a study that examines the impact of Kanban as part of a longitudinal study of communication and collaboration in software teams. Our longitudinal study investigates team communication and collaboration in a larger, distributed cloud-based setting. The research question we pose is as follows:

**Research Question:** How does Kanban impact software team's communication and collaboration?

We used the well-established framework from Porras and Hoffer for understanding team behaviors [4]. It was also used by Abrahamsson [5] to study commitment in software process improvement. Our questionnaire design was developed based on earlier work of Porras and Hoffer [4] and Abrahamssson [5]. In the next subsection, we describe the research setting in which we conducted our study.

The Software Factory is an experimental laboratory that provides an environment for research and education in software engineering, and that was established by the Department of Computer Science at the University of Helsinki [10]. We refer to our case project as Emobile. The Software Factory team developed a new prototype for the customer called BookIt. BookIt provides enterprises and organizations with user-friendly mobile solutions that work on all mobile phones and networks. The prototype integrated BookIt's established technology and connections to payment aggregators to provide a new payment option for online service providers. The project included eight engineers as well as a coach from the Software Factory. During the seven-week project, the team conducted six iterations of software development.

The Kanban process as described in [3] was followed. The team formed the columns (i.e., workflow phases for the tasks) on the Kanban board at the beginning of the project. Later on, the team altered the columns and set up WIP limits in order to get a smoother workflow. The tasks were typically chunked into ca. half-day size, i.e., three or four hours. During the 7-week project, the team had one-week iterations with weekly customer demos and retrospectives. Such a sprint-based, iterative Kanban is called Scrumban [9] and it provides a structured schedule also for the customer. Daily stand-up meetings (similar to the short, max. 15 minutes project status meetings in Scrum) were held, as well. The customers were experienced technically and in the customer role.

## II. Research Method

We used a questionnaire to collect data from the software team under study. The questionnaire included closed as well as open-ended questions. The software team had no obligation to complete the questionnaire. We requested that the team complete the same questionnaire after the end of each iteration. The repeated use of the questionnaire gave us an opportunity to examine impact patterns over the project period.

We analyzed the collected data by tracking various patterns regarding how the impact of Kanban on perceived importance and frequency of team communication and collaboration evolved during the project. We developed two versions: one with all respondents and a second with consistent respondents (i.e., those who filled in the questionnaire in at least three out of six iterations). We also analyzed responses to two open-ended questions and examined whether they were reflected in our statistical analysis. Key results from our analysis are presented in the next section.

The questionnaire focused on two themes: communicating openly and collaboration. "Communicating openly" was defined (based on Porras and Hoffer [4]) as behaviors reflecting the direct giving and receiving of information relevant to accomplishing a specific task. "Collaboration" was defined as behaviors promoting or reflecting the involvement of relevant persons in the process in identifying and solving problems. For both themes, we collected Likert scale-based inputs. The full questionnaire is available online at https://elomake.helsinki.fi/lomakkeet/34545/lomake.html.

### III. RESULTS

In this section, we briefly present the key results from the study. The questionnaire responses were analyzed based on normalized and individual scores. The responses from two open-ended questions were also reviewed.

*A. Kanban and Team Communication*

Figures 1, 2, and 3 reflect the overall perceived importance and frequency of team communication over six iterations. Figures 1 and 2 show normalized responses from all respondents in six iterations. Some respondents did not fill in the survey after each iteration. Figure 2 suggests that deviation across team members is wide in terms of how frequent they communicate with each other using Kanban as a common platform. Figure 2, however, indicates that, in general, Kanban helped the team to use communication as an important driver throughout the project, with the exception of sporadic turbulence in team communication.

Figure 3 shows the normalized score of team communication (both frequency and importance) when responses were summed together to form an overall variable. The location of the circle on the y-axis shows the frequency score, while the size of the circle and the number in it show the importance score. The overall frequency and importance of communication seem to have had consistent impacts based on Kanban, with the exception of sporadic contradictions as noted in individual responses.

The overall results show that the perceived importance of communication was a bit higher than its frequency. Kanban had the highest impact on team communication during the initial iterations. The impact, however, started to lessen when the team did not require dependence on the Kanban board to facilitate communication. One respondent said:

*"Since we have started to know each other the communication is more direct and verbal. Kanban is used but it is not the critical part on our communication. Kanban gives some structure to tasks and (when used properly) informs whats going on. However I feel that the team is not relying to Kanban board but rather ask and communicate verbally."*

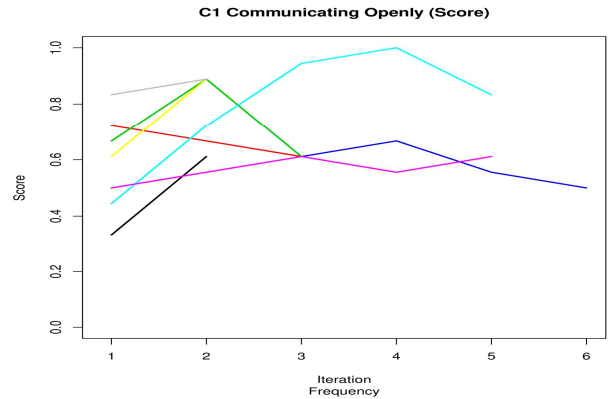

Fig. 1. The impact of Kanban on the frequency of team communication.

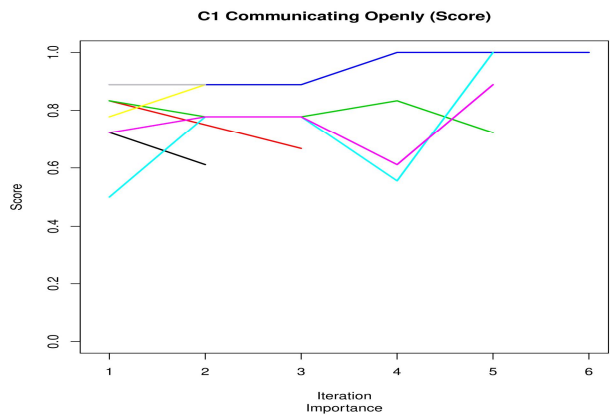

Fig. 2. The impact of Kanban on the importance of team communication.

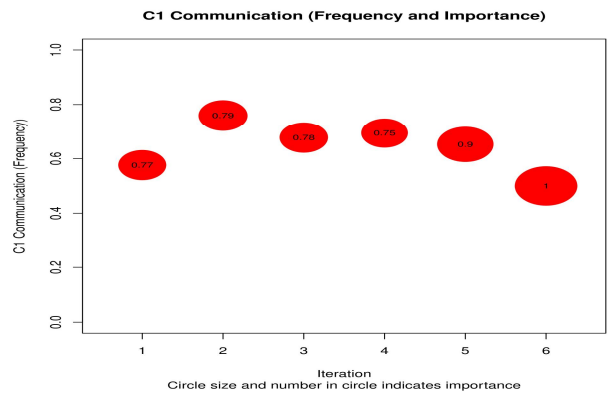

Fig. 3. The overall impact of kanban on team communication.

Based on the overall analysis, the following propositions were derived:

**Proposition 1: When team members begin to know each other and as a project matures, Kanban board may not be critical to facilitating team communication.**

*B. Kanban and Team Collaboration*

Figures 4 and 5 show that Kanban consistently helped the team remain united during the project. In particular, it helped team members to help each other as issues surfaced on the Kanban board. One respondent observed:

*"With Kanban you are aware of what other people are doing and you can always help them or monitor their work. For example, if somebody has noticed that one of the tasks is been doing for a long time, he/she can go and help that person to complete the task. It also helps solving the most important tasks first, which is crucial for the project work."*

Furthermore, the questionnaire data suggest that team collaboration took a major "nosedive" during the middle of the project (between iterations 3 and 5). Figures 4 and 5 show the frequency and importance of specific behaviors in team collaboration. It is interesting to note the light blue line in Figure 4 which starts nose-diving near the mid-point of iteration 3, then rises in the middle of the project, and then drops dramatically in the last two iterations.

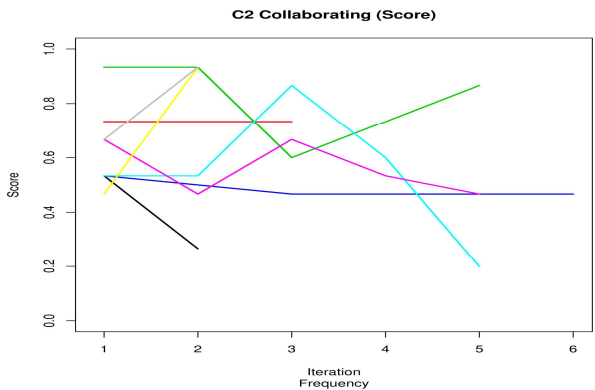

Fig. 4. The impact of Kanban on the frequency of team collaboration.

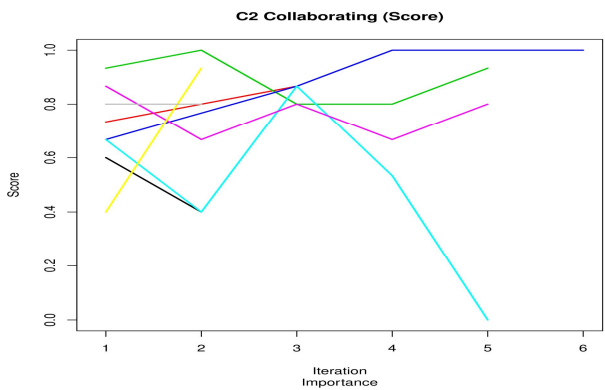

Fig. 5. The impact of Kanban on the importance of team collaboration.

Figures 4 and 5 show that the team did not follow a progressive curve regarding collaboration. It seems to have started with a varying degree of perceived importance and frequency, leading to highs and lows of collaboration throughout the project.

The main impact of Kanban may have been in keeping the core baseline collaboration going since the team had to ultimately rely on the Kanban board to advance the project.

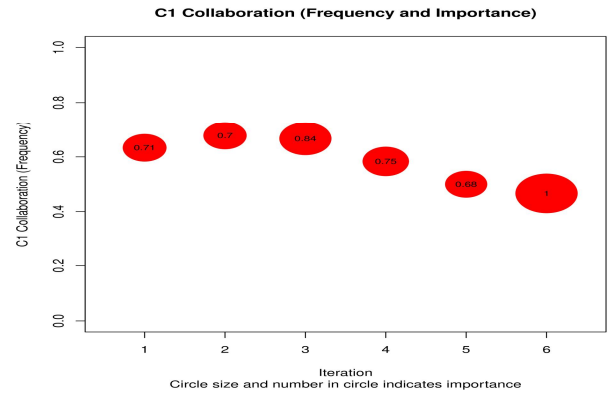

Fig. 6. The overall impact of Kanban on team collaboration.

Figure 6 shows the normalized score of team collaboration (both regarding frequency and importance). It suggests that in iteration 3, both the perceived importance and frequency of collaboration was at peak level. It then plummeted in iterations 4 and 5.

Based on overall analysis, the following proposition was derived:

**Proposition 2: Kanban may improve productivity as it encourages team members to help each other finish pending work.**

*C. Correlation Between Team Communication and Collaboration*

We also examined the relationship between team communication and collaboration. Figure 7 shows the score correlations visually in two ways. The upper right triangle of the matrix shows the data points in each combination of categories (communication – c1 and collaboration – c2, with frequencies and importance) cross-plotted. The bottom left triangle of the graph matrix shows an elliptic representation of the correlation between the variables and a LOESS smoothed curve of the same. A rounder ellipse indicates a weaker correlation. Figure 7 indicates that there is a strong correlation between C1 (frequency) and C2 (importance); that is, the frequency of collaborative behavior relates strongly to the importance of collaborative behavior. However, the data set is too small to warrant any claims of applicability to other cases. Statistically, C1 and C2 seem independent.

## IV. Discussion

Results from the study provide initial insight into the impact of Kanban on team communication and collaboration in the described context. The Kanban board may lose its impact when the team starts relying more on interpersonal communication. The study further suggests that team communication may not be related to collaboration. This may mean that Kanban acts as an interface for the team to collaborate even without strong interpersonal communication within the team.

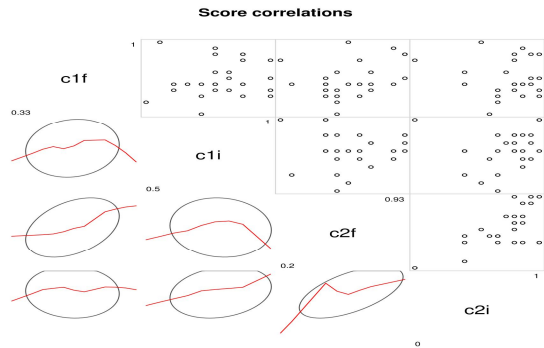

Fig. 7. The correlation between team communication and collaboration.

The consistent pattern we see in the perceived importance and frequency of collaboration is close to what Ikonen et al. [3] found in their investigation. The "nosedive" in team collaboration between iterations 3 and 5 also indicates that the sporadic breakdown in collaboration may be offset by continuous calibration of overall team efforts; it may not be due to the impact of Kanban or a specific development process in use (also reflected in [5]).

*A. Limitations*

The results from our study may not be applicable to other settings and teams using Kanban in software development. We also recognize the limitation of subjectivity in the collected qualitative data and relatively small sample size. However, we countered this limitation by using a repetitive questionnaire, allowing us to observe patterns in the data throughout the project. The results should be considered as propositions rather than affirmative or tested hypotheses. Further empirical tests in software team settings are necessary to validate the findings.

*B. Future Work*

We have set up an environment (i.e., the Software Factory) that allows longitudinal research in cooperation with the industry and other academic partners. A Software Factory network has already expanded with nodes in Madrid, Oulu, and Bolzano; in addition, more are currently being planned in Asia and North America. We have already conducted the next round of study with global teams working across the Finland and Madrid sites. We aim to further accelerate our efforts in contributing to advancing the global distributed software development body of knowledge; especially continuing on the inquiry of cloud-based distributed software teams.

V. CONCLUDING REMARKS

The presented questionnaire study serves as a starting point for a longitudinal effort to better understand means (such as Kanban) that support team communication and collaboration. The study provided initial insights on the impact of Kanban in a collocated development setting and also contributed to overall research development in team behaviors. When considering the relevance of informal communication in collocated development, we are especially interested in means that support team communication and collaborations in distributed cloud-based development environments where such informal communication is significantly hindered or absent. Based on the results of this study, we are interested in analyzing if Kanban has a consistent or progressive impact on team communication and collaboration in distributed environments by partly substituting the lack of informal communication

ACKNOWLEDGEMENTS

The study has been supported by Tekes as part of the SCABO project, the Cloud Software Program, and the Cloud Software Factory project.